\documentclass[aps,prc,twocolumn,superscriptaddress,nofootinbib,preprintnumbers]{revtex4-2}

\usepackage{graphicx}
\usepackage{dcolumn}
\usepackage{bm}

\usepackage{multirow}%
\usepackage{amsmath,amssymb,amsfonts}%
\usepackage{amsthm}%
\usepackage{mathrsfs}%
\usepackage[title]{appendix}%
\usepackage{xcolor}%
\usepackage{textcomp}%
\usepackage{booktabs}%
\usepackage{algorithm}%
\usepackage{algorithmicx}%
\usepackage{algpseudocode}%
\usepackage{listings}%
\usepackage[separate-uncertainty=true]{siunitx}
\usepackage{verbatim}
\usepackage{placeins}
 \usepackage{lineno}
\usepackage{xspace}
\usepackage{nicefrac}
\usepackage{orcidlink}
\usepackage{ulem}

\newcommand{\cawo}{CaWO$_4$\xspace}
\newcommand{\alo}{Al$_2$O$_3$\xspace}
\newcommand{\CEvNS}{CE$\nu$NS\xspace}
\newcommand{\CRAB}{\textsc{Crab}\xspace}
\newcommand{\NUCLEUS}{\textsc{Nucleus}\xspace}
\newcommand{\CRESST}{\textsc{Cresst}\xspace}
\newcommand{\FIFRELIN}{\textsc{Fifrelin}\xspace}
\newcommand{\FIFRADINA}{\textsc{Fifradina}\xspace}
\newcommand{\IRADINA}{\textsc{Iradina}\xspace}
\newcommand{\TOUCANS}{\textsc{Toucans}\xspace}
\newcommand{\GEANT}{\textsc{Geant4}\xspace}

\newcommand{\getTUWaffiliation}{\affiliation{TU Wien, Atominstitut, 1020 Wien, Austria}}
\newcommand{\getTUMaffiliation}{\affiliation{Physik-Department, TUM School of Natural Sciences, Technische Universit\"at M\"unchen, D-85747 Garching, Germany}}
\newcommand{\getMBIaffiliation}{\affiliation{Marietta Blau Institute for Particle Physics of the Austrian Academy of Sciences, A-1010 Wien, Austria}}
\newcommand{\getIRFUaffiliation}{\affiliation{IRFU, CEA, Universit\'e Paris-Saclay, 91191 Gif-sur-Yvette, France}}
\newcommand{\getMPIKaffiliation}{\affiliation{Max-Planck-Institut f\"ur Kernphysik, 69117 Heidelberg, Germany}}
\newcommand{\getINFNAaffiliation}{\affiliation{INFN, Sezione di Roma, I-00185, Roma, Italy}}
\newcommand{\getINFNBaffiliation}{\affiliation{INFN, Sezione di Roma “Tor Vergata”, I-00133 Roma, Italy}}
\newcommand{\getVERGATAaffiliation}{\affiliation{Dipartimento di Fisica, Universit\`{a} di Roma "Tor Vergata", I-00133 Roma, Italy}}
\newcommand{\getSRMPaffiliation}{\affiliation{Service de recherche en Corrosion et Comportement des Materiaux, SRMP, Universit\'e Paris-Saclay, CEA, 91191 Gif Sur Yvette, France}}
\newcommand{\getSAPIENZAaffiliation}{\affiliation{Dipartimento di Fisica, Sapienza Universit\`{a} di Roma, I-00185 Roma, Italy}}
\newcommand{\getIRESNEaffiliation}{\affiliation{CEA, DES, IRESNE, DER, Cadarache, 13108 Saint-Paul-Lez-Durance, France}}
\newcommand{\getIJCLABaffiliation}{\affiliation{Universit\'{e} Paris-Saclay, CNRS/IN2P3, IJCLab, 91405 Orsay, France}}

\begin{document}

\preprint{APS/123-QED}

\title{Observation of two nuclear recoil peaks induced by neutron capture on \alo}

\author{H.~Abele} \getTUWaffiliation
\author{P.~Ajello} \getTUMaffiliation
\author{B.~Arnold} \getMBIaffiliation
\author{E.~Bossio} \getIRFUaffiliation
\author{J.~Burkhart} \getMBIaffiliation
\author{F.~Cappella} \getINFNAaffiliation
\author{N.~Casali} \getINFNAaffiliation
\author{R.~Cerulli} \getINFNBaffiliation \getVERGATAaffiliation
\author{J-P.~Crocombette} \getSRMPaffiliation
\author{G.~del~Castello} \getINFNAaffiliation \getSAPIENZAaffiliation
\author{M.~del~Gallo~Roccagiovine} \getINFNAaffiliation \getSAPIENZAaffiliation
\author{P.~de~Marcillac} \getIJCLABaffiliation
\author{S.~Dorer} \getTUWaffiliation
\author{C.~Doutre} \email{corentin.doutre@cea.fr} \getIRFUaffiliation 
\author{A.~Erhart} \email{andreas.erhart@tum.de} \getTUMaffiliation 
\author{S.~Fichtinger} \getMBIaffiliation
\author{M.~Friedl} \getMBIaffiliation
\author{C.~Goupy}  \getIRFUaffiliation \getMPIKaffiliation
\author{D.~Hauff} \getTUMaffiliation
\author{E.~Jericha} \getTUWaffiliation
\author{M.~Kaznacheeva} \getTUMaffiliation
\author{H.~Kluck} \getMBIaffiliation
\author{T.~Lasserre}  \getIRFUaffiliation \getTUMaffiliation \getMPIKaffiliation
\author{D.~Lhuillier} \getIRFUaffiliation
\author{O.~Litaize} \getIRESNEaffiliation
\author{S.~Marnieros} \getIJCLABaffiliation
\author{R.~Martin} \getIRFUaffiliation
\author{E.~Namuth} \getTUMaffiliation
\author{T.~Ortmann} \getTUMaffiliation
\author{L.~Peters} \email[]{lilly.peter@tum.de} \getIRFUaffiliation \getTUMaffiliation \getMPIKaffiliation
\author{D.V.~Poda} \getIJCLABaffiliation
\author{F.~Reindl} \getTUWaffiliation \getMBIaffiliation
\author{W.~Reindl} \getMBIaffiliation
\author{J.~Rothe} \getTUMaffiliation
\author{N.~Schermer} \getTUMaffiliation
\author{J.~Schieck} \getTUWaffiliation \getMBIaffiliation
\author{S.~Sch\"{o}nert} \getTUMaffiliation
\author{C.~Schwertner} \getTUWaffiliation \getMBIaffiliation
\author{G.~Soum-Sidikov} \getIRFUaffiliation
\author{R.~Strauss} \getTUMaffiliation
\author{R.~Thalmeier} \getMBIaffiliation
\author{L.~Thulliez} \getIRFUaffiliation
\author{M.~Vignati} \getINFNAaffiliation \getSAPIENZAaffiliation
\author{M.~Vivier} \getIRFUaffiliation
\author{P.~Wasser} \getTUMaffiliation
\author{A.~Wex} \getTUMaffiliation

\collaboration{\CRAB Collaboration}\noaffiliation


\date{\today}

\begin{abstract}
We report the observation of two nuclear recoil peaks induced by neutron capture on aluminum in a cryogenic Al$_2$O$_3$ detector developed by the \NUCLEUS collaboration for the detection of reactor neutrinos via coherent elastic neutrino-nucleus (\CEvNS) process.
Data collected at the Technical University of Munich in 2024 with a $^{252}$Cf source reveal a main recoil line at 1145 eV from single-$\gamma$ de-excitation of $^{28}$Al and a newly observed structure near 575 eV originating from several two-$\gamma$ cascades. 
The latter constitutes the first direct measurement of a nuclear recoil line induced by multi-$\gamma$ cascades. It is predicted by our simulations when the recoiling nucleus has time to stop before the emission of the next $\gamma$-ray in the cascade. These results demonstrate the potentia performance of the \CRAB (Calibration Recoil for Accurate Bolometry) method for in situ nuclear recoil calibration and highlight the importance of accurately modeling recoil stopping and nuclear de-excitation times in cryogenic detectors of \CEvNS and dark matter interactions.
\end{abstract}

\maketitle


\section{Introduction}

Measurement of nuclear recoils in the sub-keV range offers a unique probe of neutrino properties at low energies \textit{via} the coherent elastic neutrino\,-\,nucleus scattering (\CEvNS)~\cite{Freedman:1973yd,Abdullah:2022zue}. They are also the main signature expected in the direct detection of dark matter, which remains one of the central open questions in physics and cosmology. In particular, light dark matter candidates in the MeV–GeV mass range~\cite{Angloher:2017sxg}, leading to low-energy recoils, have not yet been fully explored and are actively searched for. Both processes produce sub-keV nuclear recoils, demanding detectors with extreme sensitivity. In recent years, significant progress has been achieved with cm-scale cryogenic calorimeters reaching eV-scale thresholds~\cite{CRESST:2019jnq,Strauss:2017cuu,PhysRevD.99.082003,EDELWEISS:2022ktt,PhysRevLett.127.061801}. A precise interpretation of \CEvNS and dark-matter data, however, requires an accurate calibration of the detector response to low-energy nuclear recoils -- beyond the standard electronic-recoil calibrations provided by X-ray or LED sources~\cite{nucleus2025xrf}.

This is precisely the aim of the \CRAB (Calibrated Recoil for Accurate Bolometry) method, which uses radiative neutron capture on the nuclei of the cryogenic detector to induce nuclear recoils of known energy. These recoils are identical in every respect to those that would be produced by neutrinos or light dark matter scattering off nuclei~\cite{Thulliez:2020esw}. The first validations of this method were achieved with the direct detection of a tungsten recoil peak at 112.5~eV in the \cawo detectors of the \NUCLEUS~\cite{PhysRevLett.130.211802} and \CRESST~\cite{PhysRevD.108.022005} collaborations, using $^{252}$Cf and Am-Be commercial neutron sources installed next to their respective cryostat. This nuclear recoil peak is expected due to the emission of a single high-energy $\gamma$-ray after neutron capture on the isotope $^{182}$W. Lately, the \CRESST experiment has reported the observation of a similar recoil peak at 1145~eV induced by neutron capture on $^{27}$Al in an \alo detector~\cite{cresst2025sapphirepeak}. 

In this paper, we present a new measurement of the nuclear recoil spectrum induced by neutron capture in a \alo cryogenic calorimeter. It was conducted with a \NUCLEUS \alo detector at the shallow Underground Laboratory of the Technical University (TU) of Munich, using a $^{252}$Cf neutron source. Thanks to high detector performance combined with relatively long measurements, we detected two nuclear recoil peaks, as predicted in~\cite{Soum2023, abele2025crabfacilitytuwien} when taking into account the timing effect of $\gamma$ de-excitations.
In Sect.~\ref{sec:setup} we describe the experimental setup and the acquired data sets. The data analysis procedure -- including data processing, the statistical framework for peak detection, and the discussion of the energy scale -- is presented in Sect.~\ref{sec:data_analysis}. Section \ref{sec:data_sim_comparison} introduces the \CRAB simulation package and compares the measured rates in the recoil peaks with their predictions. Finally, Sect.~\ref{sec:conclusion} outlines future prospects for high-precision measurements with \alo with the dedicated \CRAB calibration facility installed near a research reactor of the TU of Vienna~\cite{abele2025crabfacilitytuwien}, benefiting from higher thermal neutron fluxes, reduced backgrounds and the possibility of cross-calibration using electron recoil lines from X-ray sources and LEDs.

\section{Experimental Setup, Data Acquisition and Data Sets}
\label{sec:setup}

The experimental setup consists of a 0.5~g cubic Al$_2$O$_3$ absorber crystal equipped with a tungsten thin-film transition-edge sensor (TES) having a normal-to-superconducting transition temperature of 18~mK. The detector builds on \CRESST TES technology \cite{GRAPES-3:2024yym} and was developed for the first phase of \NUCLEUS \cite{nucleus2025commissioning}. The absorber crystal is enclosed in a $\simeq$1\,cm thick copper support box that shields the detector against thermal radiation from warmer stages and is held in place by a flexible bronze clamp with point-like contacts realized via 1~mm sapphire spheres. A thin gold wire provides the thermal connection between the crystal and the copper holder. During the measurement, the TES was biased at a current of $I_\mathrm{B} = 2.4\,\mu\mathrm{A}$ and actively stabilized using an integrated resistive heater on the crystal surface. A heater control pulse was injected every 50 s to monitor the stability of the detector response. For electronic recoil calibration, the detector was continuously exposed to a non-collimated $^{55}$Fe source mounted in the lid of the detector holder, providing the characteristic Mn K$_\alpha$ line at 5.9~keV and the K$_\beta$ line at 6.5~keV. 

The detector was operated in a wet dilution refrigerator (Kelvinox\,100, Oxford Instruments plc) designated for the \CRAB calibration facility, prior to its relocation to the TRIGA~Mark~II reactor site at TU~Vienna~\cite{abele2025crabfacilitytuwien}. 
At the time of the measurement, the dilution refrigerator was located at TU~Munich and surrounded by a 14~cm thick cylindrical lead shielding to reduce environmental radioactivity. The system reaches a base temperature below 10\,mK and requires weekly liquid helium refilling.

\begin{figure}
    \centering
    \includegraphics[width=1\linewidth]{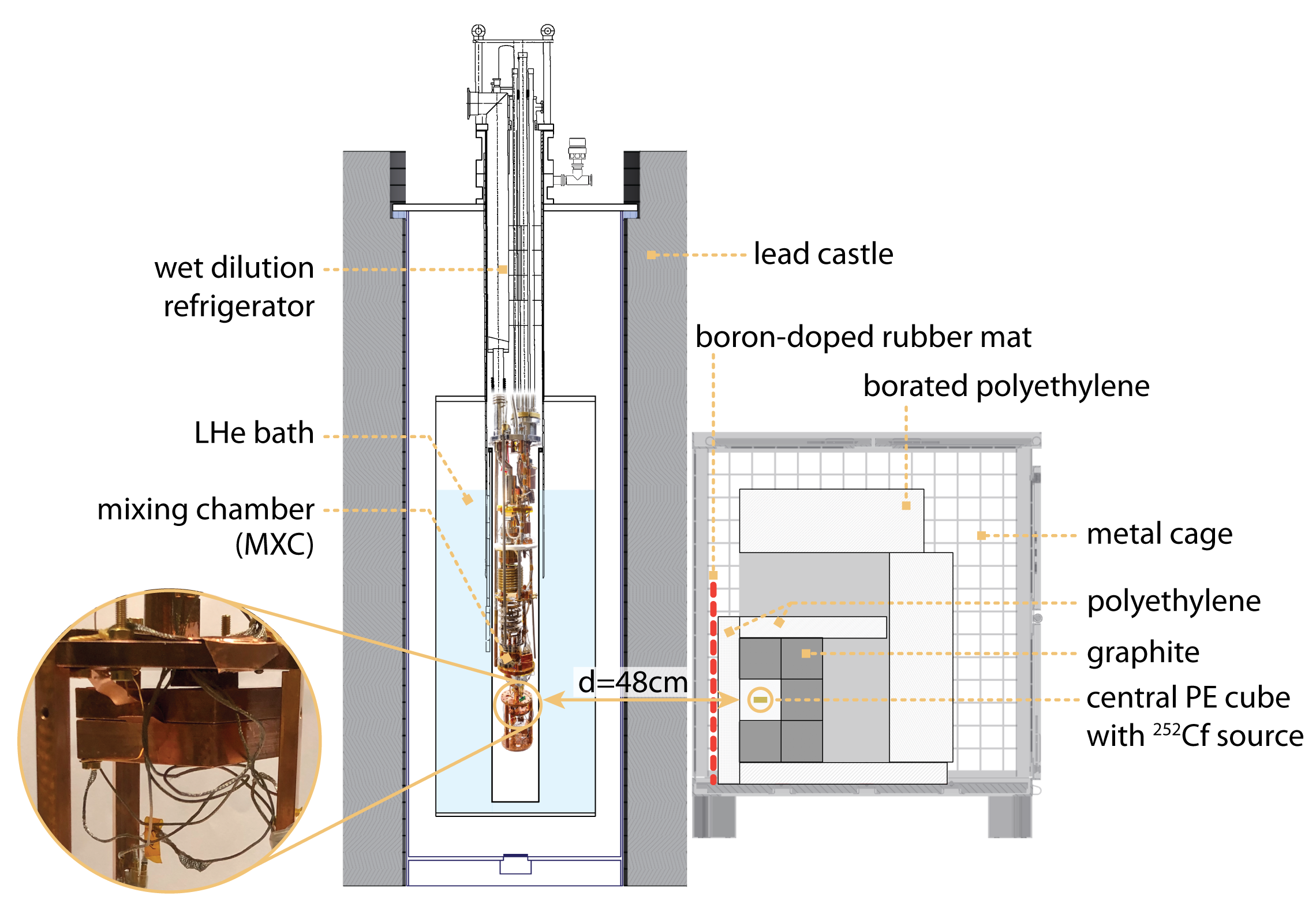}
    \caption{Schematic of the experimental setup. The dilution cryostat is surrounded by 14 cm thick lead shielding, against which the source and its shielding are placed. A 5\,mm thick boron-loaded rubber mat ($\mathrm{B_4C}$) can be inserted between the source and the lead shielding to block the thermal neutrons. The inset shows the copper housing of the cryogenic detector.}
    \label{fig:exp_setup}
\end{figure}

An illustration of the experimental arrangement is shown in Fig.~\ref{fig:exp_setup}. The setup was exposed to a flux of thermal neutrons originating from a commercial $^{252}$Cf source with an activity of 2.6~MBq (at the time of measurement), installed in a compact moderator setup developed in~\cite{PhysRevLett.130.211802}. To thermalize the fast neutrons from the source (average energy of 2.12~MeV), the source capsule was placed at the center of a polyethylene (PE) cube with 10\,cm edges, with an additional 5\,cm thick PE layer towards the cryostat. Graphite blocks placed on the sides and back of the central PE cube act as additional moderators and reflectors, enhancing the forward-directed neutron flux. For more details, refer to \cite{PhysRevLett.130.211802}. The assembly was positioned against the lead castle of the cryostat, reducing the flux of $\gamma$-rays from the $^{252}$Cf source. For radiation protection, a layer of borated PE was installed to reduce the ambient dose rate, and the entire setup was enclosed in a locked metal cage for radio-protection purpose.

The detector signals were amplified and read out using a DC-SQUID system (Magnicon) and digitized with an Incaa Computers VD80455 transient recorder at a sampling frequency of 200~kHz. A hardware trigger threshold of 3.4~mV was applied corresponding to about 24 eV according to the calibration coefficient determined in Section~\ref{sec:data_analysis}. For each triggered event a record window of 8192 samples, equivalent to 40.96~ms, was stored. The trigger time was set to the first quarter of this window, allowing the baseline level to be determined before the pulse and restored before the end of the window.

Measurements were carried out from February~20 to April~8,~2024, within a single cryogenic run. Three data sets were acquired. First, background data were collected over a total measuring time of 373.2~h to optimize detector performance and characterize the ambient background rate. Subsequently, the neutron source setup was installed outside the cryostat and source data were recorded over 346.2~h, with the source positioned at a distance of \SI{48(1)}{\cm} from the cryogenic detector. Finally, an additional 158.5~h of data was collected with a 5~mm boron carbide-doped rubber mat ($\mathrm{B_4C}$) inserted between the neutron source and the lead castle. The mat efficiently captures thermal neutrons from the source while leaving the fast-neutron and $\gamma$ backgrounds largely unaffected, thus providing a reference measurement for source-induced background.

\section{Data Analysis}
\label{sec:data_analysis}

To analyze the data, two different analysis frameworks were used: \textsc{CryoLab}~\cite{nucleus2025xrf}, a MATLAB-based software developed for the analysis of \NUCLEUS and \CRAB data, and \textsc{Cait}, a Python3 package for processing raw data from cryogenic detectors~\cite{Wagner_2022}. The two analyses yield compatible results and \textsc{CryoLab} is taken as the reference in the following.

As described in Sect.~\ref{sec:setup}, the triggering is performed online during the data acquisition. The measurement exhibited excellent stability and a low baseline constant noise. This was verified through regular injections of artificial heater pulses and by monitoring, the physical event rate, the noise power spectrum and the baseline resolution over time, which remained constants. However, we observed that the trigger efficiency was reaching a flat plateau only after $\approx$450 eV amplitude, far above the hardware trigger setpoint. The origin of this high actual value of trigger threshold is not understood.

Two classes of pulses are observed in the data. Slow pulses occur at a constant rate, show a dominant thermal component, and are attributed to mechanical signal transmission from the Al$_2$O$_3$ support spheres rather than to particle interactions in the crystal. In contrast, fast pulses are correlated with the particle rate and are interpreted as genuine energy depositions in the crystal bulk.
To suppress artifacts and distorted signals, we apply mild pulse-shape cuts that efficiently remove the slow events while leaving the fast, physical pulses unaffected. We verified that these cuts do not introduce any sizable energy dependence in the detection efficiency above 450\,eV.
The data were acquired with a hardware-triggered system, so the continuous stream was not available. As a consequence, the absolute value of the efficiency plateau is difficult to determine with precision. The analysis presented here therefore focuses on identifying the structures in the recoil spectrum and determining their relative intensities.
Pulse amplitudes are reconstructed using two complementary methods. First, an optimal filter~\cite{Gatti:1986cw}, constructed from the measured noise-power spectrum and a pulse template, provides the most precise energy estimates in the low-energy region. Second, a truncated pulse-shape fit is used to recover the energy of saturated events. Both estimators are calibrated independently and cross-checked using external calibration sources and reconstructed spectra.

To derive a pulse template, the thermal model~\cite{Probst:1995hjq} is fitted to the signals and the fit parameters are averaged  yielding a rise time of $\tau_{\text{ath}} = 0.04$\,ms, decay times of $\tau_{\text{in}} = 0.06$\,ms and $\tau_{\text{th}} = 2.2$\,ms. The ratio of the amplitudes of the athermal to thermal signals is set at 0.8.

The baseline resolution is estimated by superimposing the pulse template on a large set of noise traces and applying the optimal filter. The distribution of reconstructed pulse amplitudes follows a Gaussian shape with its standard deviation yielding an estimate for the baseline resolution of \SI{0.589(14)}{\milli\volt}, equivalent to an excellent energy resolution of \SI{4.1(0.1)}{\eV} based on the 6~keV X-ray calibration described below.
\begin{figure}
    \centering
    \includegraphics[width=\linewidth]{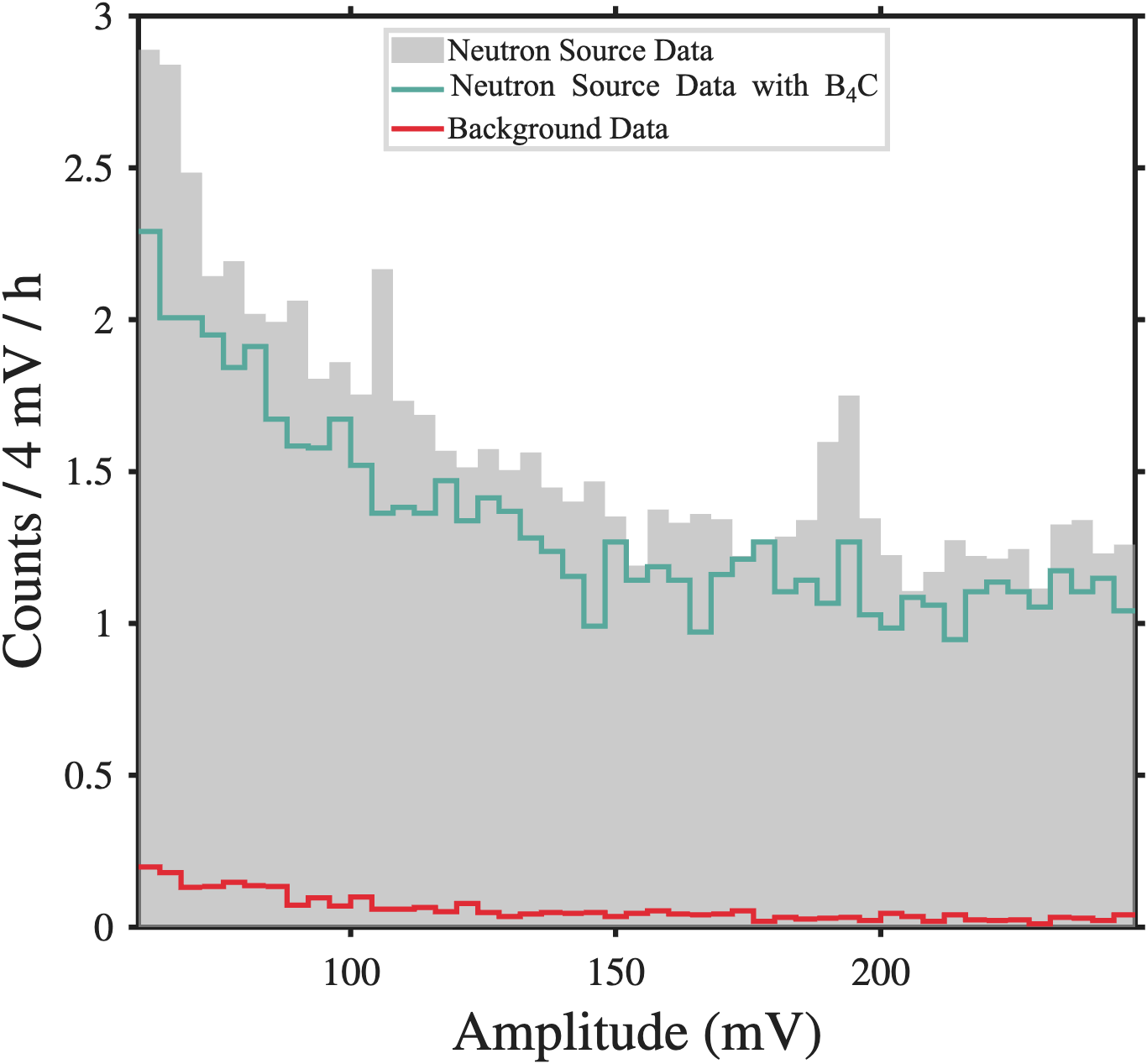}
    \caption{\label{fig:spectrum} Energy spectra in voltage units of the neutron source (gray shaded histogram) and background data (red) in the region of interest. For the neutron source data, an additional data set with a boron carbide shielding ($\mathrm{B_4C}$), blocking all thermal neutrons but only a small fraction of fast neutrons, is shown (green).}
\end{figure}
Figure~\ref{fig:spectrum} shows the reconstructed energy spectra in voltage units for the different data sets described in Sect.~\ref{sec:setup} up to 250\,mV (1.75 keV), a region above which pulses saturate. The truncated pulse-shape fit technique allows to extend energy reconstruction to higher levels, and yields the pulse height spectrum displayed in Fig.~\ref{fig:spectrum_iron}.
The Mn K$_\alpha$ line is visible in all data sets, but the K$_\beta$ line is hidden in the data with neutron source due to an enhanced background. The calibration based on the position of the K$_\alpha$ line leads to a linear calibration constant of 7.0~eV/mV. 
In addition to the Mn lines, the Cu K$_\alpha$ line is present in the background and neutron source data without B$_4$C. Copper is the most abundant material near the detector and the comparison of the different spectra in Fig.~\ref{fig:spectrum_iron} shows that, in the configuration of our experiment, its fluorescence lines are mainly activated by neutron capture and, to a lesser extent, by cosmic muons~\cite{nucleus2025commissioning}. 

\begin{figure}
    \centering
    \includegraphics[width=\linewidth]{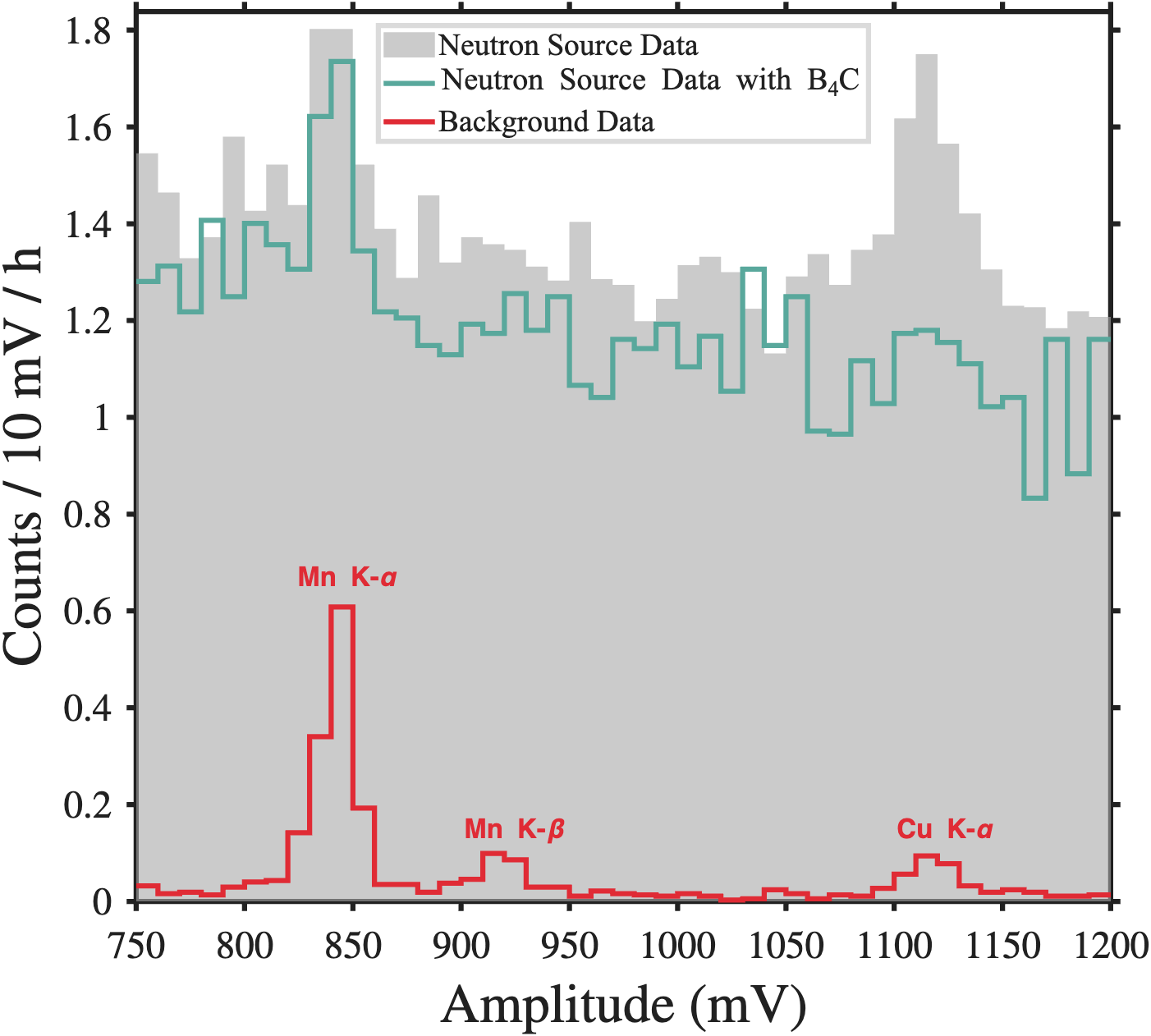}
    \caption{\label{fig:spectrum_iron} Pulse height spectra reconstructed with the truncated pulse shape fit showing the Mn K$_\alpha$(5.89 keV), Mn K$_\beta$ (6.49 keV) and Cu K$_\alpha$ (8.04 keV) lines in the neutron source and background data sets. The color code is the same as in Figure~\ref{fig:spectrum}.}
\end{figure}

\subsection{Low energy blind peak search}
\label{subsec:main-peak}

A preliminary analysis performed on the neutron source data revealed two apparent peaks in the 500–1500~eV range (Fig.~\ref{fig:spectrum}), based on an initial linear calibration with the Mn K$_\alpha$ line but without a quantitative estimate of significance. For a more reliable and unbiased result, a blind peak search was later performed, ensuring objective identification of spectral features and consistent determination of their positions and significances.
The neutron source spectrum of Fig.~\ref{fig:spectrum} is scanned progressively with a 60~mV-wide moving window. In each window, the local background is empirically described by the sum of an exponential and a constant term, and the tested peak feature is modeled by a Gaussian function, whose initial guess for its position and width are the center of the window and the baseline resolution respectively. Two $\chi^{2}$ minimization's -- \textit{background only} and \textit{signal + background} -- are then performed with the background model parameters and the mean position of the peak left free, whereas its amplitude and width are forced to be positive. A statistical test of the null hypothesis (\textit{background only}) is then defined as $\Delta\chi^{2} = \chi^{2}_\text{bkgd+signal} - \chi^{2}_\text{bkgd}$. The distribution of $\Delta\chi^{2}$ is obtained by generating numerous statistical realizations of the background model and found to follow a $\chi^{2}$ law with an effective number of Degrees of Freedom (DoF) of $3.35 \pm 0.05$, slightly deviating from the expectation of 3 DoF. This fitted $\chi^{2}$ law is then used to determine the one-sided significance (in $\sigma$ unit) for rejecting the \textit{background only} hypothesis. This approach is more conservative than the use of the default 3~DoF. 

\begin{figure}
    \centering
        \includegraphics[width=\linewidth]{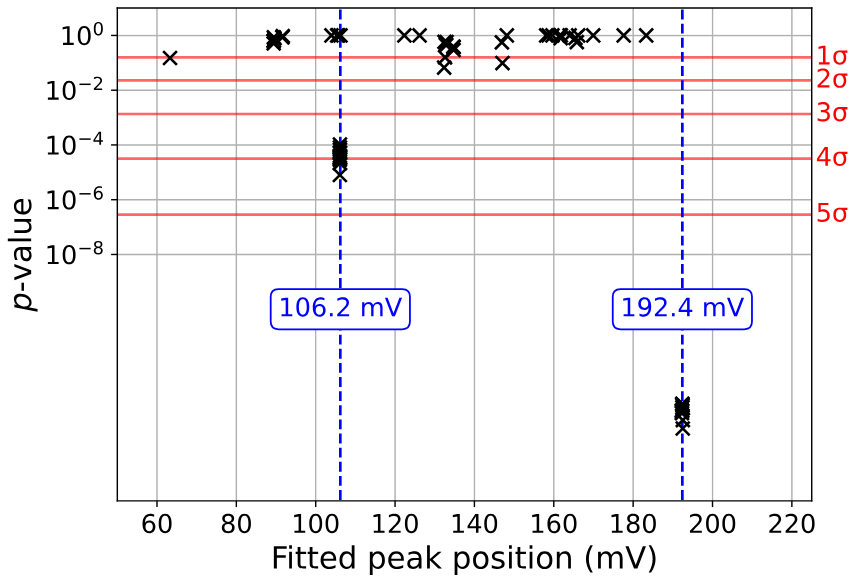}
    \caption{\label{fig:blind_peak_search} Results of the blind peak search applied to the neutron source spectrum of Fig.~\ref{fig:spectrum}. Each point corresponds to a given position in the scan of the 60~mV-wide window. Only two peaks, with mean position around 106~mV and 192~mV (blue dashed lines), are clearly identified. When the sliding window contains one of these two positions, the null hypothesis (\textit{background only}) has a p-value $\le10^{-4}$ (left vertical axis) or equivalently a rejection significance $\ge 4\sigma$ (right vertical axis).}
\end{figure}

\begin{figure*}
    \centering
    \includegraphics[width=\linewidth]{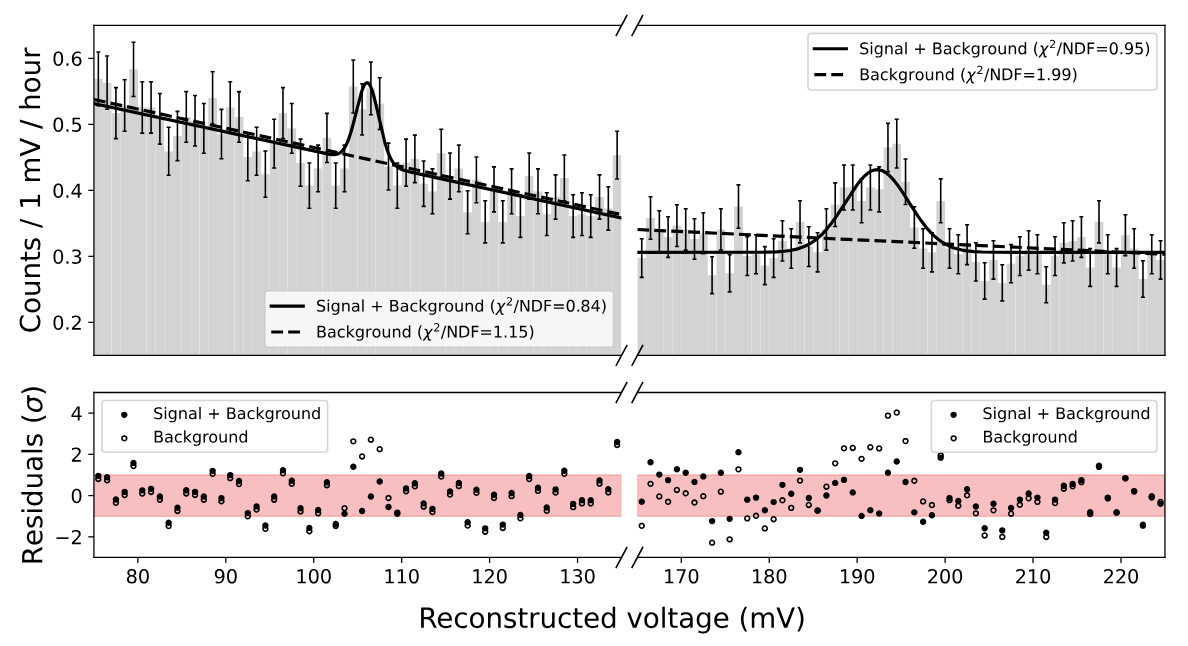}
    \caption{Top row: pulse height spectra in voltage units of the neutron source data after background subtraction, showing the low (left) and high energy (right) peaks identified with the blind peak search. The source related background is modeled as an exponential function plus a constant. The solid and dashed lines show the best fits with and without a Gaussian signal, respectively. The obtained count rates contained in the Gaussian peaks are \SI{11.6(0.9)} and \SI{28.7(1.4)} counts per day. The bottom row shows the fit residuals, compatible with a normal distribution in the \textit{signal+background} hypothesis.}
    \label{fig:fit_peaks}
\end{figure*}

Figure~\ref{fig:blind_peak_search} shows that the blind search identifies only two peak-like features in the experimental recoil spectrum with significance $\ge$ 4$\sigma$. In order to determine whether these peaks can be interpreted as the nuclear recoil lines expected from neutron capture (see Sect.~\ref{sec:data_sim_comparison}), we performed local fits around the two pointed mean positions. The results of these fits are illustrated in Fig.~\ref{fig:fit_peaks}. To assess systematic uncertainty in the mean positions, the background model, bin width, and fitting range have been independently varied. The obtained best fit mean positions are then $106.2 \pm 0.3\text{ (stat)} \pm 0.1\text{ (syst)}$~mV and $192.4 \pm 0.5\text{ (stat)} \pm 0.1\text{ (syst)}$~mV. The one-sided significance of the rejection of the null hypothesis is 4.0$\sigma$ and 7.6$\sigma$ respectively. These fits also allow us to compute the ratio of the rates in the first peak to that in the second peak, $0.40~\pm~0.03~\text{ (stat)}~\pm~0.07\text{ (syst)}$, which will be compared to a prediction in Sect.~\ref{sec:data_sim_comparison}.

\subsection{Discussion of Energy Scale}
\label{subsec:calibration-discussion}

The two recoil peaks detected above background are clearly induced by thermal neutrons. As can be seen from Fig.~\ref{fig:spectrum}, they disappear when the boron carbide shielding is in place, strengthening the hypothesis of \CRAB peaks induced by neutron capture on $^{27}$Al: the peak at highest energy would be induced by the single-$\gamma$ transitions with 1145 eV nuclear recoil energy and the second peak would result from timing effects in several 2-$\gamma$ cascades with a mean recoil energy of 575 eV~\cite{Soum2023, abele2025crabfacilitytuwien}. Considering the linear calibration constant of 7.0~eV/mV, obtained from the position of the Mn K$_\alpha$, the reconstructed energy of the 2-$\gamma$ and single-$\gamma$ peaks are $743.4 \pm 2.8\text{ (stat)} \pm 0.7\text{ (syst)}$~eV and  $1346.8 \pm 4.2\text{ (stat)} \pm 1.4\text{ (syst)}$~eV, corresponding to a relative discrepancy with respect to a prediction of 29.2\% and 17.6\% respectively. It is expected that the electronic and nuclear energy scales will differ due to the different energy loss processes involved, affecting the phonon signal detected. However, according to recent simulations \cite{PhysRevD.106.063012} and measurement by the \CRESST experiment \cite{cresst2025sapphirepeak}, the expected deviation in \alo is within the few percent range, much lower than the values observed here. Furthermore, the fact that the relative differences depend strongly on the energy hints towards a significant detector non-linearity.

A X-ray fluorescence (XRF) source allows to study the linearity of the detector response to electron recoils from the Mn K$_\alpha$ line down to sub-keV energies. In~\cite{nucleus2025xrf}, a significant detector non-linearity has been observed using this technique with a similar \NUCLEUS \cawo detector. The \alo detector used in this work has also been exposed to a slightly modified XRF source. The measurement is described in more detail in the appendix. It shows a strong non-linearity of the detector response, yielding a deviation of up to 45\,\% at 1\,keV from the linear Mn K$_\alpha$ calibration (Fig.~\ref{fig:xrf-calcurve}). It is noteworthy that the detector behavior and the strength of the non-linearity may vary from run to run depending on the TES operating point on the transition curve between superconducting-normal states. Still, this XRF measurement shows a clear indication that a significant deviation from linearity as suggested by the position of the nuclear recoil peaks is plausible.

\section{Discussion and comparison with simulation}
\label{sec:data_sim_comparison}
The simulation of the setup was performed with the \TOUCANS code~\cite{THULLIEZ2023}, based on \GEANT~\cite{GEANT4,Geant42016}, using the NeutronHP package to transport low-energy neutrons ($\leq$20 MeV). This software package now matches the performance of reference neutron transport codes~\cite{Thulliez2022,Zmeskal2023,Zmeskal2024}. The neutron source simulation starts with by sampling $^{252}$Cf fission neutron and $\gamma$ events produced by the \FIFRELIN code~\cite{Litaize2015,Litaize2023} before propagating all particles in the source shielding; for more details refer to~\cite{PhysRevLett.130.211802}. The $^{252}$Cf activity is used to normalize the equivalent time of the simulation. The cryostat geometry is described in~\cite{abele2025crabfacilitytuwien}. 
Here, the significant amount of lead and PE between the source and the detector (see Fig.~\ref{fig:exp_setup}) makes the simulation very computationally demanding since the probability for one neutron from the source to contribute to the calibration peak at 1145 eV is found to be of the order of 10$^{-10}$. For this reason, the simulated measurement time is only around half of the experimental live time. 

\begin{figure}
    \centering
    \includegraphics[width=1\linewidth]{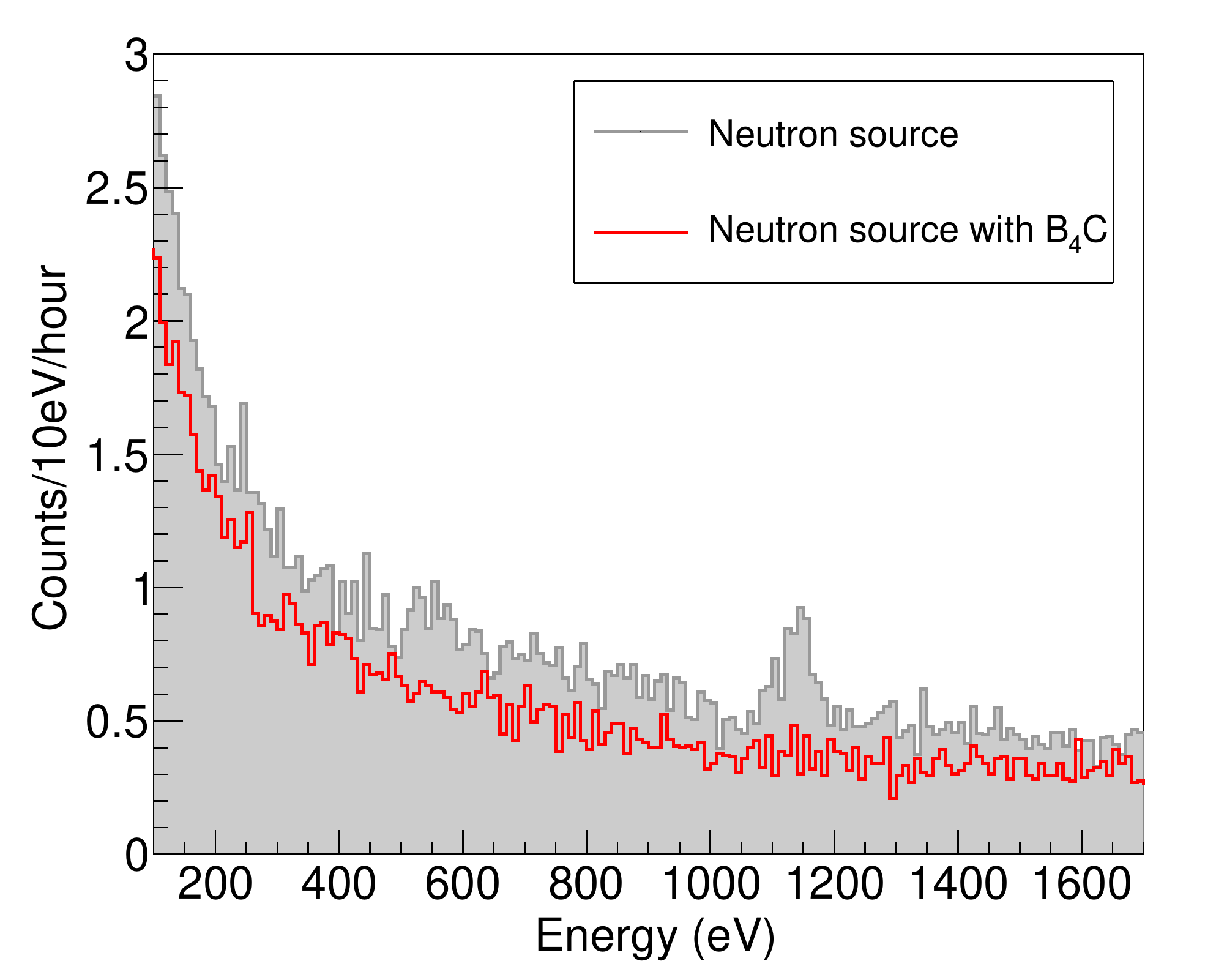}
    \caption{Predicted energy spectra induced by the neutron source, without (black, 192~h equivalent time) and with (red, 153~h equivalent time) the boron carbide screen. The experimental resolution is used to convolve the simulation results.}
    \label{fig:simu_spectra}
\end{figure}

When a neutron capture occurs in the cryogenic detector, the induced nuclear recoil and associated $\gamma$-rays are sampled from the output of the \FIFRADINA code~\cite{Soum2023}. This package couples the \FIFRELIN code, which simulates nuclear de-excitation, with the \IRADINA code \cite{Borschel2011} for atom transport in matter. It thus accounts for the interplay between the time evolution of the $\gamma$ cascade and collision cascade. Propagation of all emitted particles from the capture reaction vertex with \GEANT leads to the predicted recoil spectra shown in Fig.~\ref{fig:simu_spectra}. Qualitatively they confirm the overall shape of the background spectrum and validate the choice of the effective \textit{exponential + constant} model used in Sect.~\ref{sec:data_analysis}. The effect of the boron carbide rubber mat is also compatible with data. 

The thermal neutron flux inside the \alo crystal is estimated to be around 10$^{-1}$~n/cm$^{2}$/s. The main recoil peak at 1145~eV is clearly visible in Fig.~\ref{fig:simu_spectra} with a predicted rate of \SI{54.4(6.3)}{counts/day}, that is a factor \SI{1.93(0.24)} higher than the experimental rate (see Fig.~\ref{fig:fit_peaks}). Part of this discrepancy comes from the detection efficiency, which is not corrected here. Also, as already observed in previous studies~\cite{PhysRevLett.130.211802}, the predicted rate of thermal neutrons at the crystal location is very sensitive to the detailed geometry around the detector. In particular, the information on the orientation of the detector housing with respect to the source has been lost in this measurement, and its impact could be sufficient to explain much of this discrepancy in normalization. The limited statistics and the resolution effect do not allow us to clearly identify other peak-like structures in the simulated spectrum that we might have expected from timing effects.

\begin{figure}
    \centering
    \includegraphics[width=1\linewidth]{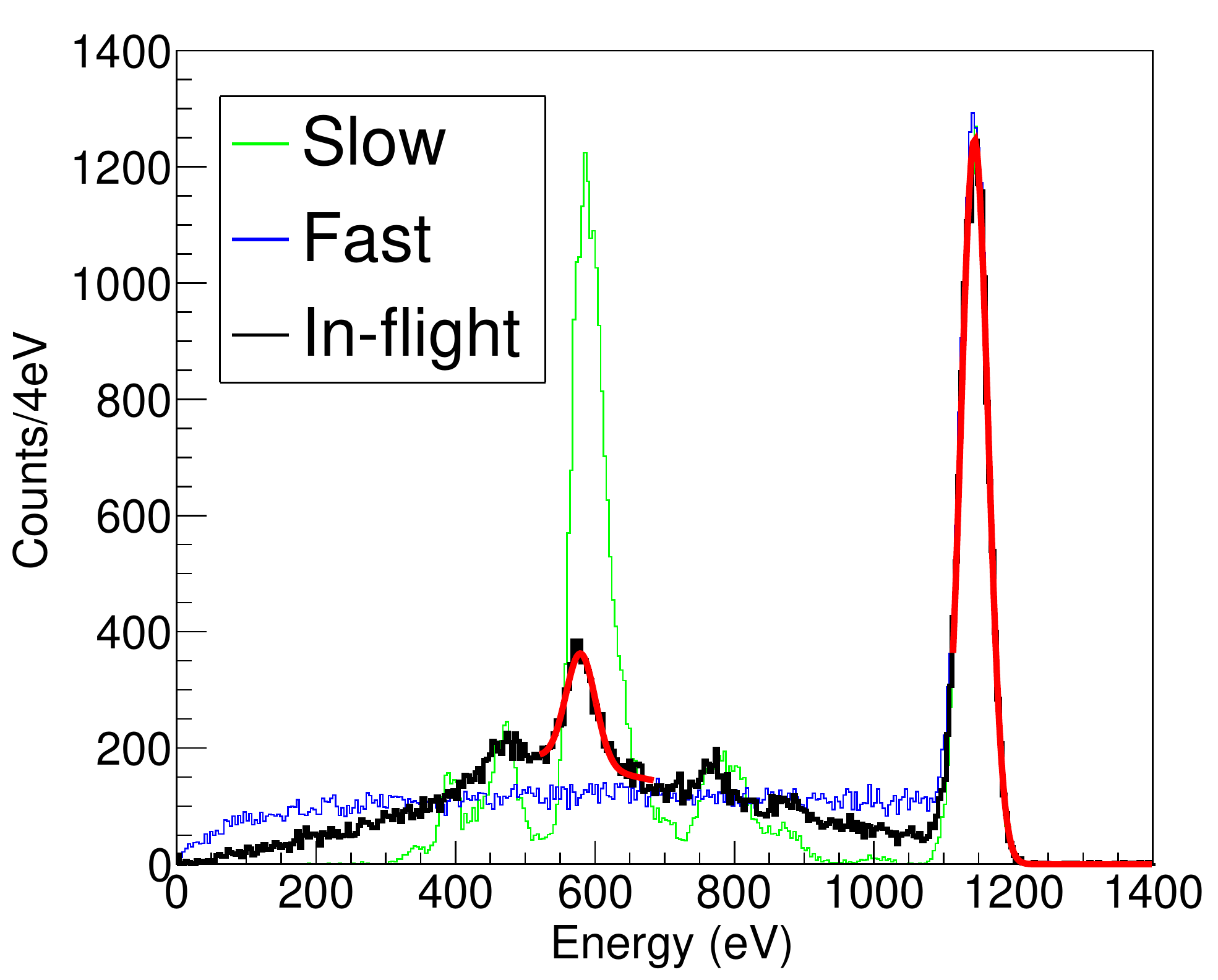}
    \caption{Al$_2$O$_3$ nuclear recoil spectra simulated with the \FIFRADINA and \GEANT package for 10$^7$ thermal neutrons incident on the cryodetector. Different timing hypotheses have a large impact on the feature at 575 eV. The "Slow" simulation (green) assumes that the recoiling nucleus has always time to stop in the detector crystal before emitting the next $\gamma$ in the de-excitation cascade. The "Fast" simulation (blue) assumes that all $\gamma$ rays in a cascade are promptly emitted and then the nucleus recoils and stops. The "In-flight" simulation (black) is our reference prediction including our best knowledge of the nuclear levels half-life and of the time course of the atomic displacement cascades. The red curves illustrate the fits of the two main recoil peaks.}
    \label{fig:simu_al2o3}
\end{figure}

Combining the limitations of the \GEANT simulation with the large non-linearity of the experimental energy scale, we do not try to adjust the predicted spectra to the experimental ones, instead we focus on the local fits of the recoil peaks and their relative intensities as presented in Sect.~\ref{sec:data_analysis}. To this end we simulated with high statistics the distribution of the pure signal by sending thermal neutrons directly to the cryogenic detector. The predicted nuclear recoil spectrum (black line), convoluted with the experimental energy resolution, is shown in Fig.~\ref{fig:simu_al2o3}. The higher statistical accuracy of this simulation and the absence of fast neutron background from the source reveals the rich structure of the recoil spectrum with, in addition to the single-$\gamma$ peak at 1145~eV, another prominent feature at 575~eV resulting from timing effects in several two-$\gamma$ cascades. Using the latest evaluated nuclear half-lives from the RIPL-3 database and the description of displacement cascades provided by \FIFRADINA, the predicted ratio of the count rates of the peak at 575~eV to that of the peak at 1145~eV is $0.17 \pm 0.01\text{ (stat)} \pm 0.01\text{ (syst)}$, where the systematic uncertainty is inferred from several fits of the Gaussian peak with different range and background shape hypotheses. This ratio is significantly smaller than the measured ratio, by a factor \SI{2.2(0.4)}..
We consider that the predicted intensity of the peak at 1145~eV is robust since it corresponds to a single-$\gamma$ transition with a known branching ratio listed in the nuclear databases. On the other hand, the amplitude of the feature at 575~eV depends on many more ingredients that are not as well known: the half-life of the intermediate nuclear levels (between 3.4 and 4.0 MeV) in seven different 2-$\gamma$ cascades and how it compares to the stopping time of the recoil nuclei in \alo. For $^{28}$Al most of the nuclear half-lives are measured and indexed in the RIPL-3 database. Although large variations, up to a factor 3, have been observed in the past between different data sets, most recent results converge to the same values. Interestingly, these half-life measurements use the in-flight emission of $\gamma$-rays by nuclei recoiling in matter (Doppler Shift Attenuation method)~\cite{Kangasmaki_1994}, which relies on Binary Collision Approximation (BCA) codes such as \IRADINA or SRIM~\cite{ZIEGLER2010} to compute the trajectories of the nuclei. In this BCA approaches the lattice structure of the material is neglected; atoms are assumed to be isotropically distributed in space, and the flight length between two collisions is sampled from a Poisson distribution centered on the mean interatomic distance. Our observation of a higher-than-expected intensity of the 575 eV peak relative to the 1145 eV peak may indicate an overestimation of nuclear stopping times by BCA codes, with potential impact on the interpretation of half-life measurements. 
To better understand this effect dedicated molecular dynamics (MD) calculations are in preparation. The flight time distributions of recoil nuclei provided by these simulations will, \textit{a priori}, be the most realistic that can be obtained.  Agreement with our average value for the intensity ratio of the two peaks would require a correction by a factor of about 3 towards shorter flight times in matter, or longer nuclear lifetimes.

\section{Conclusion and Outlook}
\label{sec:conclusion}

The measurement presented demonstrates a successful application of the \CRAB method to an Al$_2$O$_3$ cryogenic detector operated within the \NUCLEUS setup at the Technical University of Munich. Exposure to thermal neutrons from a moderated $^{252}$Cf source produced two distinct peak features in the recoil energy spectrum, consistent with expectation from neutron capture on $^{27}$Al. A statistical analysis confirmed the presence of the low-energy peak with a 4.0$\sigma$ significance and 7.6$\sigma$ for the high-energy peak. The relative accuracy in the determination of the mean position of each nuclear recoil line is 0.4\%. This underlines the potential of the \CRAB method to provide accurate anchor points of the nuclear recoil energy scale. 

The main recoil peak, expected at 1145\,eV, arises from single-$\gamma$ de-excitation of the compound nucleus, and the secondary structure, expected at 575 eV, originates from timing effects in two-$\gamma$ cascades. While the main peak has been recently observed by the \CRESST collaboration~\cite{cresst2025sapphirepeak}, the secondary peak represents the first indication of two-$\gamma$ cascade signatures, offering a novel experimental probe of the distribution of recoil stopping times and of nuclear level lifetimes in the few-femtosecond range. This validation of new calibration features arising from timing effects in multi-$\gamma$ de-excitation cascades also strengthens the interest in \CRAB measurements using other detector materials, such as germanium~\cite{abele2025crabfacilitytuwien}.

 Using the electronic recoil calibration from the Mn K$_\alpha$ X-ray line, the energy of the two nuclear recoil peaks are reconstructed at $743.4 \pm 2.8\text{ (stat)} \pm 0.7\text{ (syst)}$~eV and $1346.8 \pm 4.2\text{ (stat)} \pm 1.4\text{ (syst)}$~eV. The large, energy-dependent discrepancies with respect to expected values cannot be explained solely by the underlying different energy loss processes at work in electron and nuclear recoils. Rather, it is consistent with strong non-linearities in our cryogenic detector, as already observed in complementary XRF measurements. 

Comparison between experimental and predicted ratio of the two peak intensities showed a factor 2.2 $\pm$ 0.4 discrepancy. This observable is a unique probe of potential biases in nuclear level half-lives or in recoil stopping times or both. Simulation work is ongoing to compare the stopping time predictions from BCA and MD codes.

In the upcoming second phase of the CRAB project at the TU Vienna reactor facility~\cite{abele2025crabfacilitytuwien}, enhanced thermal neutron fluxes and reduced backgrounds will allow high precision measurements of the nuclear recoil energy scale. Valuable insights to further improve the understanding of the detector response are anticipated from simultaneous measurements with an XRF source. The Cu L$_\alpha$ and Al K$_\alpha$ X-ray lines at 923 and 1486\,eV are located near the single-$\gamma$ transition nuclear recoil line at 1145\,eV, enabling a direct comparison between electronic and nuclear recoils around 1 keV with little sensitivity to intrinsic non-linearity of the detector. A similar approach can be taken by comparing the response to nuclear recoils at 575 eV with that to 677 eV X-rays from fluorine. The X-ray energy depositions at the surface will be complemented by optical calibrations with an LED providing energy depositions distributed throughout the crystal bulk. Together, these complementary calibration techniques will establish a precise validation of the detector response at sub-keV energies, a critical step for future \CEvNS and light dark matter measurements.

\begin{acknowledgements}
We acknowledge funding by DFG through the SFB 1258 and the Excellence Cluster ORIGINS, by the European Commission through the ERC-StG2018-804228 “NUCLEUS”, and by the Austrian Science Fund (FWF) through the projects “I 5427-N CRAB” and “P 34778-N ELOISE”. 

\end{acknowledgements}

\bibliography{crab-ref}

\clearpage 
\FloatBarrier
\section*{Appendix}

\begin{figure}
    \centering
    \includegraphics[width=\linewidth]{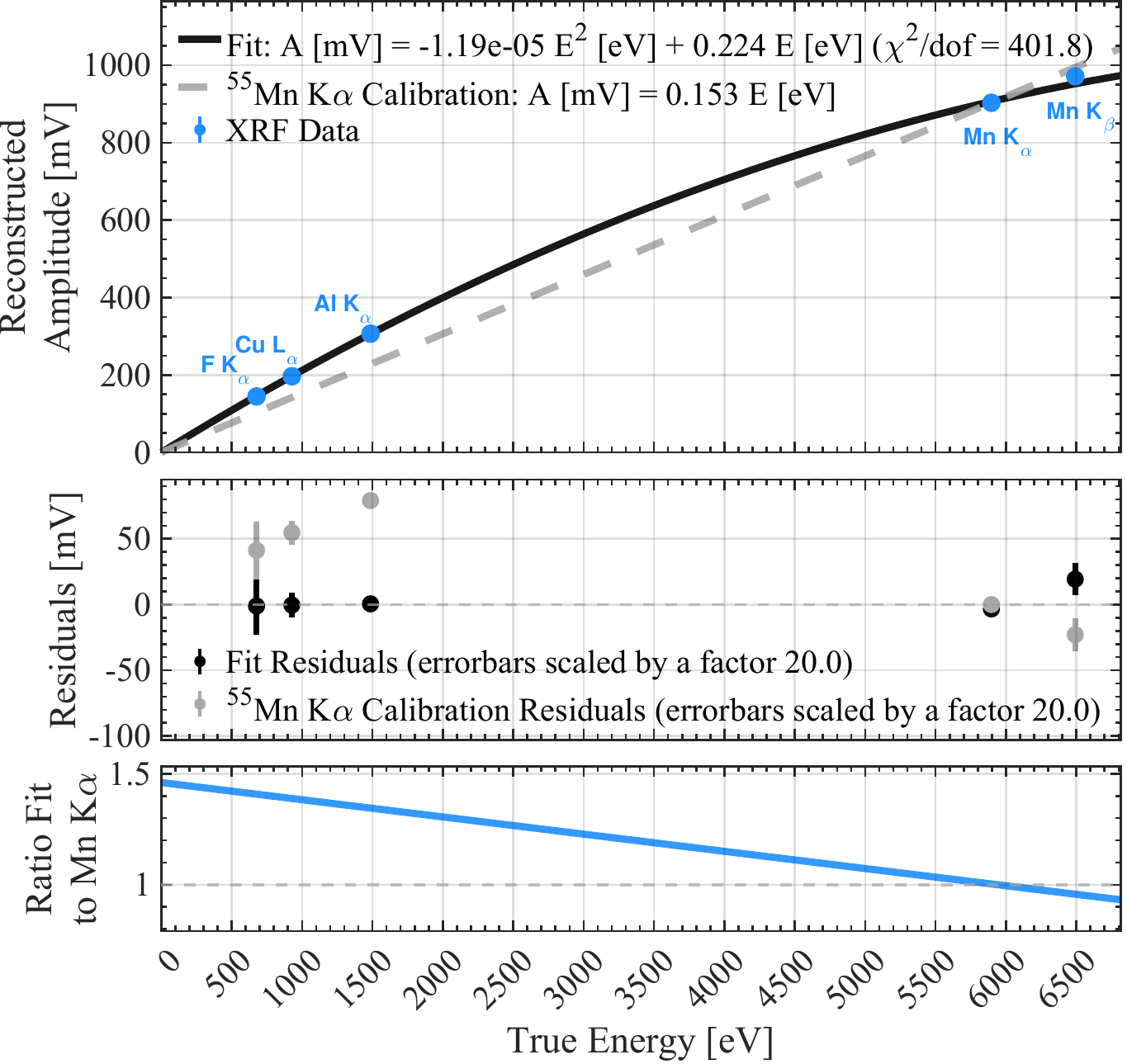}
    \caption{\label{fig:xrf-calcurve}Top panel: Reconstructed amplitudes (in mV) of the identified X-ray lines from the XRF measurement, plotted as a function of true energy (in eV). The blue points represent the measured data. Two calibration models are shown: a linear interpolation anchored at zero and the Mn K$_\alpha$ line (gray dashed line), and a quadratic fit to all data points (black solid line). 
	Middle panel: Residuals between the measured data and the model predictions, plotted as a function of true energy. Residuals from the linear fit are shown in gray, and those from the quadratic fit in black. Here, the uncertainties have been rescaled so that the reduced chi-square is normalized to unity.
	Bottom panel: Ratio of the quadratic fit to the linear calibration curve, highlighting the systematic error in the energy scale that would result from relying on the linear model anchored at the Mn K$_\alpha$ line.}
\end{figure}

\begin{figure}
    \centering
    \includegraphics[width=\linewidth]{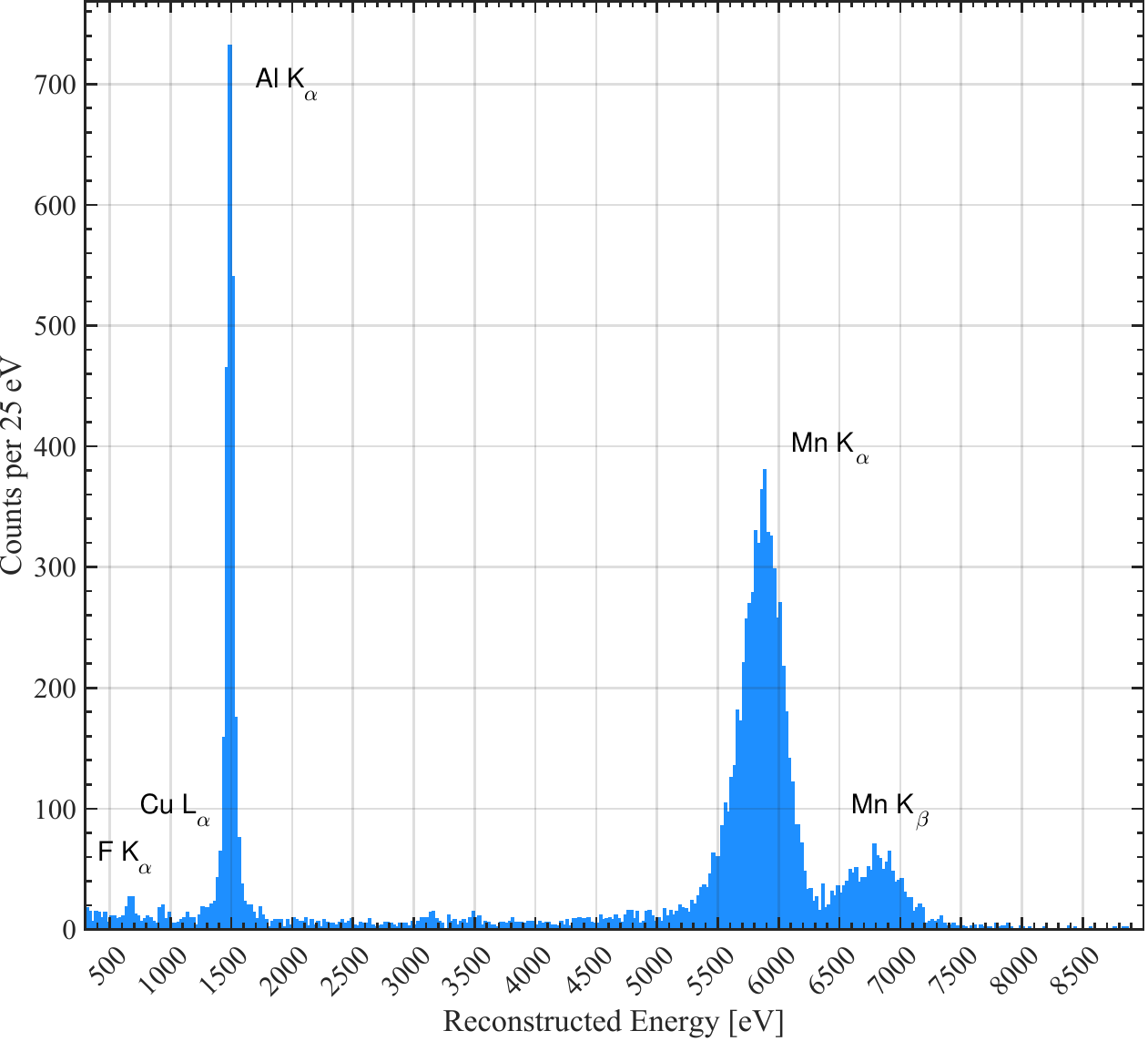}
    \caption{\label{fig:xrf-spectrum}Calibrated energy spectrum, using the quadratic calibration shown in Figure \ref{fig:xrf-calcurve}. The elemental transitions corresponding to the most prominent peaks are indicated.}
\end{figure}

In the following, we briefly present a measurement conducted in a dry dilution refrigerator in an above-ground laboratory at the physics department of the Technical University of Munich, where the \alo detector used in this work was exposed to an XRF source. The approach is similar to what was presented in~\cite{nucleus2025xrf}, employing an initial $^{55}$Fe source irradiating a two-staged target arrangement. However, some aspects of the source design were modified with the goal to decrease the measurement time necessary to observe a calibration line at sub-keV energies. The primary radiation from the $^{55}$Fe source (\SI{9.81e5}{\becquerel}) illuminates the first target consisting of a squared aluminum piece (approx. $\SI{22}{mm}\times\SI{22}{mm}$). The second target is formed of a PTFE piece (approx. $\SI{25}{mm}\times\SI{22}{mm}$). Both targets were equipped with longitudinal triangular grooves, to reduce the probability of elastic scattering and to increase the emission surface area.

The measurement lasted for about 50\,h. The signals are fully saturated in a large part of the spectrum, allowing only the truncated pulse shape fit reconstruction to be used for linearity studies. The most prominent peaks in the measured spectrum correspond to \(\mathrm{F}\) K$_\alpha$ (0.677\,keV),   \(\mathrm{Al}\) K$_\alpha$ (1.486\,keV), \(\mathrm{Mn}\) K$_\alpha$ (5.897\,keV) and \(\mathrm{Mn}\) K$_\beta$ (6.492\,keV), also an indication for the \(\mathrm{Cu}\) L$_\alpha$ (0.927\,keV) line is visible. The \(\mathrm{Cu}\) K$_\alpha$ and K$_\beta$ lines cannot be activated because their energy is higher than the primary radiation from the $^{55}$Fe source and the exposure time is too short to observe the activation by cosmics rays.
Figure \ref{fig:xrf-calcurve} shows the resulting calibration curve, displaying the reconstructed amplitude of the identified X-ray lines as a function of true energy. As in~\cite{nucleus2025xrf}, the quadratic model yields a significantly better description of the data than the linear interpolation of the \(\mathrm{Mn}\) K$_\alpha$ calibration. The deviation from the linear calibration reaches a maximum of \(45\%\) at baseline.
The measured energy spectrum is shown in Fig.~\ref{fig:xrf-spectrum}, adopting the quadratic model for calibration.

\end{document}